\documentclass[aps, pre, reprint]{revtex4-1}
\pdfoutput=1
\usepackage{amsmath}
\usepackage{graphicx}
\usepackage{url}

\begin{document}
\title{Heterogeneous contact networks in COVID-19 spreading: the role of social deprivation}

\author{Arnab~Majumdar}
\affiliation{ArcVision~Technologies, 34 Mahatma Gandhi Road, Kolkata 700009, India}

\author{Anita~Mehta}
\email[Corresponding author: ]{anita.mehta@ling-phil.ox.ac.uk}
\affiliation{Centre for Linguistics and Philology, Walton Street, Oxford OX1 2HG, UK}
\affiliation{Max Planck Institute for Mathematics in the Sciences, Inselstra{\ss}e 22, 04103 Leipzig, Germany}

\begin{abstract}
  We have two main aims in this paper. First we use theories of disease spreading on networks
  to look at the COVID-19 epidemic on the basis of individual contacts --- these give rise to
  predictions which are often rather different from the homogeneous mixing approaches usually
  used. Our second aim is to look at the role of social deprivation, again using networks as
  our basis, in the spread of this epidemic. We choose the city of Kolkata as a case study,
  but assert that the insights so obtained are applicable to a wide variety of urban environments
  which are densely populated and where social inequalities are rampant. Our predictions of
  hotspots are found to be in good agreement with those currently being identified empirically
  as containment zones and provide a useful guide for identifying potential areas of concern.
\end{abstract}


\maketitle

\section{Introduction}

The global crisis caused by the onset of the novel Coronavirus (COVID-19) pandemic has caused a flurry of academic activity across many disciplines, ranging from epidemiology to statistical physics. Most ongoing statistical physics research has been geared towards getting concrete answers in terms of infected populations, deaths, and preventive measures as quickly and simply as possible. This is, of course, an extremely useful approach to take, given that policy-makers need simple models to craft broad and easily understandable solutions. In our opinion, however, a balance needs to be struck between simplicity and accuracy in the interests of efficiency of outcome. It is for this reason that we focus on an approach that is both more rigorous and more intuitive, which asserts that it is not sufficient to treat this problem within the homogeneous mixing approaches so far used by physicists; we need in fact to focus on the contact networks of individuals, so that we can account for the stark difference in impact of those who have a larger number of contacts and are much more likely to contract the infection and subsequently propagate the disease, than those who live relatively isolated existences.

The Susceptible-Infected-Removed (SIR) model is one of the first to have been used for disease propagation \cite{bailey-1975, anderson-1991}, and consists of a population that is Susceptible, some of whom can be Infected, while others are Removed (recover or die). This has been widely used in the current pandemic and has given rise to the popular use of the parameter $R_0$, which is the number of individuals that are on average infected by a person who is infected. The simple idea behind this is that for values of this parameter greater than 1, the disease spreads and will eventually become an epidemic (at a rate proportional to $R_0$), while if the number of people infected by an infected person is kept well below 1, the epidemic will die out.

This takes no account of the fact that such a model assumes that the parameter $R_0$ is universal across a population, i.e. everyone has an equal likelihood of transmitting the infection to the same number of people. This is based on an assumption that the population is homogeneous and well mixed, and that everyone is capable of transmitting the infection to everyone else. Such an ``on average'' assumption, however, fails dangerously in the case of epidemics. Newman~\cite{newman-2002} was the first to take into account that individuals needed to be resolved in terms of their `degree distribution', i.e. the number of people that they were in contact with, and his pioneering solutions to the disease propagation network have since been widely used~\cite{meyers-2005,pastor-satorras-2015,meyers-2006,meyers-2007} for epidemics ranging from HIV to SARS-1. In the following section (Materials and Methods), we review some of the formalism that is relevant to our modelling.

One of the important fallouts of the networks approach is the natural occurrence of hotspots, i.e. of regions of high connectivity, which are particularly vulnerable to the spread of disease. Given the lack of widely available contact network data for such situations, we focus on the conditions of strict lockdown and postulate that the size of an individual's contact network is primarily determined by the size of their household. Regions where household sizes are large are often those which are socially deprived, e.g. when low-income groups live in cramped conditions. These poorer areas provide flashpoints for the propagation of disease even in cities where most of the population live in more privileged conditions. We use available data on the city of Kolkata in India as an example of a city with a high population density which contains many areas of social deprivation, and show how both factors contribute strongly to epidemic propagation. We begin with an analysis of the city as a whole, specialising next to the wards (local areas) that comprise it, so that the heterogeneities of contact networks are probed on smaller scales; we find that this more microscopic analysis has the effect of enhancing outbreaks, and speeding up transitions to epidemics locally. This tendency is even further amplified when we extend our analysis to the population of slum dwellers, where social deprivation adds to cramped living conditions to create even larger contact networks among people who are forced to access basic facilities together. One of the consequences of the above analysis is that we can outline a geographical map of hotspots where quarantining and testing should, in fact, be focused; our predictions are in good agreement with empirical estimates by the government and will, we hope, provide guidelines for future planning in the case of areas not yet identified empirically.

\section{Materials and Methods}

\subsection{Model of contact networks}

We first provide a brief review of Newman's seminal work~\cite{newman-2002} on disease propagation on networks. Individuals on a network are linked by disease-causing contacts, which Newman uses to define the probability of transmission of the disease as the transmissibility $T$, in terms of which relevant quantities are defined. The contact network is defined by the number of contacts or `degree' $k_i$ that an individual $i$ is in contact with. The degree $k_i$ is a random variable drawn from a degree distribution $p_k$. Here we follow Newman in considering a network with degree distribution defined by
\begin{equation}
  p_k = C k^{-\alpha} e^{-k/\kappa} \quad \mathrm{for} \ k \ge 1
\end{equation}
where $C$ is a normalizing constant, $\alpha$ is the power-law exponent, and $\kappa$ defines the cut-off. Such a distribution is both flexible in expressing various real-world networks as well as stable~\cite{newman-2002}. The importance of this in our case is to do with the fact that we would expect a lot of our degree distributions to have long tails, which are characterised by power-law distributions; these will be critically important for the eventual evolution to epidemics. However, power-law distributions are an idealisation to infinite systems and are, in reality, cut off by exponential tails~\cite{newman-2002}.

Newman~\cite{newman-2002} obtained closed analytical expressions for various entities based on this form of $p_k$.  The normalization constant $C$, mean connectivity $\langle k \rangle$ and the mean squared connectivity $\langle k^2 \rangle$ can be computed as
\begin{subequations}
  \begin{eqnarray}
    C &=& \frac{1}{\mathrm{Li}_{\alpha}\left[e^{-1/\kappa}\right]}\\
    \langle k \rangle &=& \frac{\mathrm{Li}_{\alpha-1}\left[e^{-1/\kappa}\right]}
                               {\mathrm{Li}_{\alpha}\left[e^{-1/\kappa}\right]}\\
    \langle k^2 \rangle &=& \frac{\mathrm{Li}_{\alpha-2}\left[e^{-1/\kappa}\right]}
                               {\mathrm{Li}_{\alpha}\left[e^{-1/\kappa}\right]}
  \end{eqnarray}
\end{subequations}
where $\mathrm{Li}_{n} \left[ x \right]$ is the $n$-th polylogarithm of $x$.

In the context of epidemics, we are concerned with the size of an infected cluster beginning with a single infected individual. By virtue of his or her connectivity, an infected individual could potentially infect a subset of $k$ connected individuals who were initially susceptible but uninfected. The probability that an infected individual transmits the infection to a connected uninfected contact is defined as the transmissibility $T$. For low values of $T$, the number of transmissions do not reach epidemic proportions so that a relatively small cluster of people is infected. Beyond a critical threshold $T_\text{c}$ a transition occurs and a significant proportion of the population are infected. The threshold $T_\text{c}$ can be computed~\cite{newman-2002,meyers-2005,meyers-2006,meyers-2007} as
\begin{eqnarray}
  \label{eq-Tc}
  T_\text{c} &=& \frac{\mathrm{Li}_{\alpha-1}\left[e^{-1/\kappa}\right]}
                      {\left(
                        \mathrm{Li}_{\alpha-2}\left[e^{-1/\kappa}\right] -
                        \mathrm{Li}_{\alpha-1}\left[e^{-1/\kappa}\right]
                       \right)} \\
               &=& \frac{\langle k \rangle}{\langle k^2 \rangle - \langle k \rangle} \nonumber
\end{eqnarray}

This is akin [8] to the percolation threshold on the underlying contact network, defined purely by topological parameters. Since $\kappa$ determines the cut-off for the power-law domain of the distribution $p_k$, the limit $k \to \infty$ corresponds to $p_k \sim k^{-\alpha}$ and the results obtained for pure power-law distributions hold. Equation~(\ref{eq-Tc}) takes the form
\begin{equation*}
  T_\text{c} \to \frac{\zeta(\alpha-1)}{\zeta(\alpha-2)- \zeta(\alpha-1)} \quad
  \text{as}\ k \to \infty,
\end{equation*}
where $\zeta(\cdot)$ is the Riemann Zeta function. As a consequence, $T_\text{c}<0$ for $\alpha<3$. However, for finite $\kappa$, $T_\text{c}$ can be finite even for these smaller values of $\alpha$.

For $T<T_\text{c}$ , the mean cluster size of infected individuals $\langle s \rangle$ can be computed as
\begin{eqnarray}
  \langle s \rangle &=& 1 +
                     \frac{
                         \left( \mathrm{Li}_{\alpha-1} \left[e^{-1/\kappa}\right] \right)^2
                        }{ \mathrm{Li}_{\alpha} \left[e^{-1/\kappa}\right]
                        } \times \nonumber\\
                    &&    \frac{T}
                            {
                               (T + 1)\ \mathrm{Li}_{\alpha-1} \left[e^{-1/\kappa}\right]
                               - T\ \mathrm{Li}_{\alpha-2} \left[e^{-1/\kappa}\right]
                            }
\end{eqnarray}

Similar to percolation, this mean cluster size diverges at the transition as
\begin{equation}
  \langle s \rangle \sim \left| T_\text{c} - T \right|^{-\gamma}
  \quad \mathrm{for}\ T \to T_\text{c}^{-}
\end{equation}
where $\gamma$ is the critical exponent.

For $T>T_\text{c}$ , the cluster of infected individuals $S(T)$ is a finite fraction of the population and can be computed by solving a self-consistent relation as shown in~\cite{newman-2002}. Again, analogous to percolation, $S(T)$ is the order parameter which grows as $S \sim \left|T-T_\text{c} \right|^\beta$ for $T \to T_\text{c}^+$. In the limit $\kappa \to \infty$, we have a pure power-law distribution where the scaling exponent $\beta$ is related to the distribution exponent $\alpha$~\cite{pastor-satorras-2015}, as
\begin{equation*}
  \beta =
    \begin{cases}
      1/\left(3 - \alpha\right)& \mathrm{for} \ \alpha < 3\\
      1/\left(\alpha - 3\right)& \mathrm{for} \ 3 < \alpha \le 4\\
      1& \mathrm{for} \ \alpha > 4
    \end{cases}
\end{equation*}

We note that the basic reproduction number $R_0$, which is the average number of people to whom an infected individual transmits the disease in homogeneous approaches, is related to the topological parameters $\alpha$ and $\kappa$, as well as the transmissibility $T$ via:
\begin{equation}
  R_0 = T\,\frac{\langle k^2 \rangle - \langle k \rangle}{\langle k \rangle}
\end{equation}
The above relationship ensures that at the epidemic threshold $T=T_\text{c}$ , $R_0=1$, as it should be~\cite{meyers-2005,meyers-2006,meyers-2007}.

\subsection{Estimating model parameters from Census data}

To estimate the parameters of our model we make use of publicly available data from the Census of India, 2011~[9]. 

We use the \texttt{HH-01CITY} dataset which contains the number of households with sizes 1, 2, 3, 4, 5, 6, 7-10, 11-14, and 15+ for the entire city of Kolkata. From this distribution, we compute the distribution of contact degree $k$. We assume that for a household of size $H$, each of the $H$ members has a degree $k=H-1$~[10]. Combining this with the number $N_H$ of households obtained from the data, we construct the cumulative distribution $P_k$. The cumulative distribution is matched to our model from Eq. (1), by fitting the parameters $\alpha$ and $\kappa$. Since $\alpha$ is connected to the fundamental structure of the contact network, we fix its value for all subsequent estimations and adjust the cut-off parameter $\kappa$.

We use the primary census abstract \texttt{DDW-PCA} which is granular to the level of the 141 wards within the Kolkata Municipal Corporation (KMC) to obtain the heterogeneity of the contact networks within the city. For this purpose, the contact network is adjusted to match the results for the mean household size for the ward, while the computation for the mean cluster-size above Tc is scaled to the population of the ward.

Some of the information about slum populations is extracted from the ``Kolkata Municipal Corporation Percentage of Slum Population to Total Population'' published for the 2011 Census of India [9]. Each ward is then segregated into ``slum'' and ``non-slum'' sub-populations. For the non-slum population, the computation in the paragraph above is retained. For the slum population, we rely on more recent data~[11] which suggests that 83\% of the households do not have in-house sanitation facilities or water supply and are thus forced to be in contact with at least one other household, thereby increasing the size of the contact network of each member living there.

\section{Results}

The network formalism is, as mentioned above, the most appropriate one to examine the transmission of COVID-19 since all transmission takes place through human contacts. Individual contact networks completely determine the spread of the infection --- infected people who live secluded existences have few, if any, people to infect in turn, while those with large familial and social networks are capable of infecting many people once they are themselves infected.

As policy-makers realise this, the importance of tools that provide data on contact networks is being increasingly realised both at the levels of academia [12] and government [13]. However, the difficulty of getting accurate data, as well as the fact that people are mobile in general and tend to infect people even without knowing them (e.g. in public places) makes this a difficult enterprise. Although efforts are currently in place to estimate global data in the way people move, at a macroscopic level (i.e. without reference to individual infected people) [14], it may be a while before accurate data on contact networks, relevant to the spread of infection, are publicly available in most democratic countries.

This inherent complexity has had the consequence that much of the discussion among scientists and policy-makers has centred around `homogeneous' SIR theory-related approaches, which are both easier to understand and which via somewhat sweeping assumptions that every individual is equal on average to every other from the point of infection-spreading, are much more tractable. While there are situations [15] where such assumptions of homogeneous mixing may well hold, there are many more situations where local and heterogeneous aspects are critical, and where predictions from homogeneous models can be somewhat misleading. 

\begin{figure}[b]
  \includegraphics[width=\columnwidth]{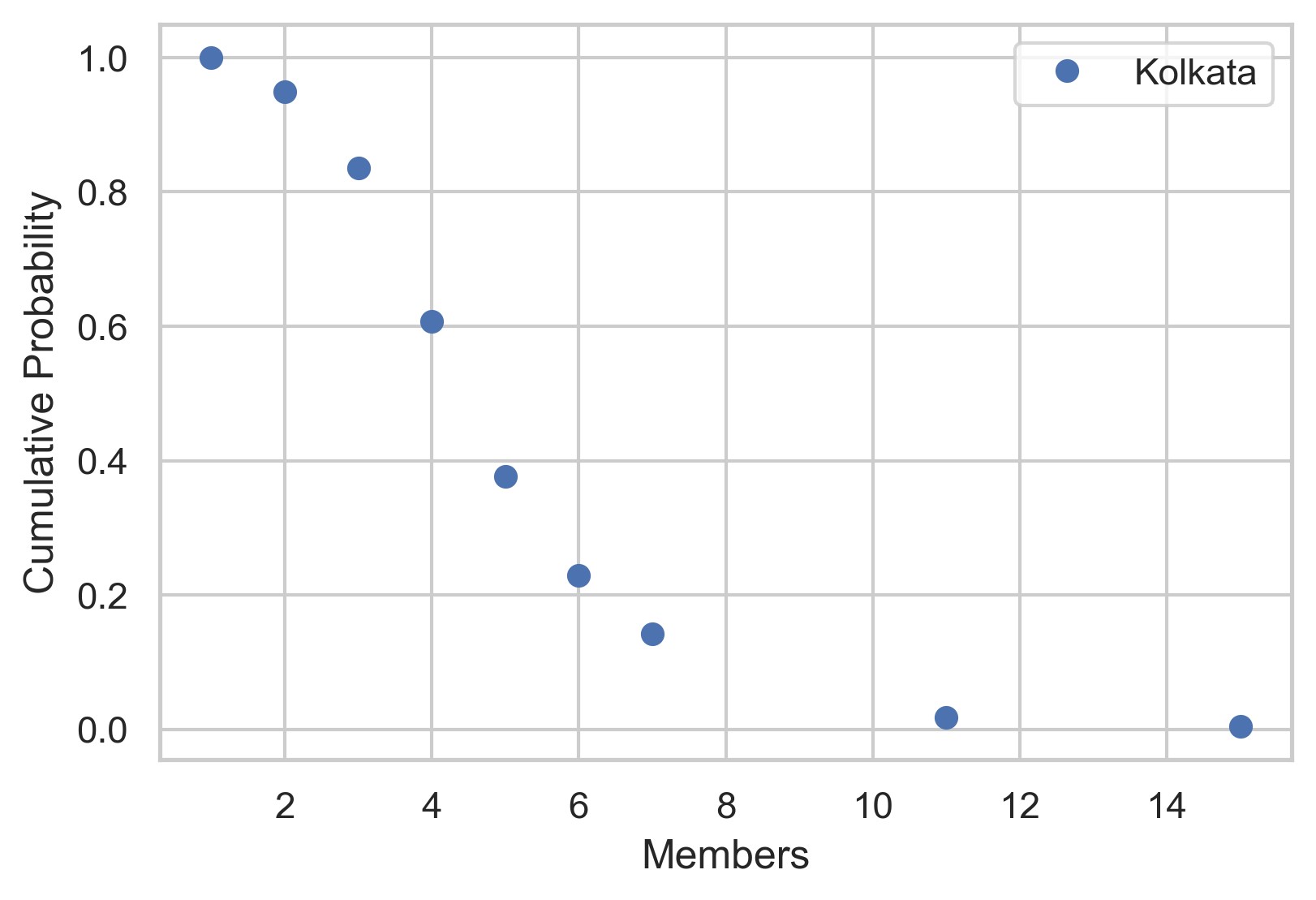}
  \caption{\label{fig-1}
  The cumulative household size distribution for the city of Kolkata from 2011 census data [9]. The total population size is 4,261,627, the number of households is 1,007,365, and the mean household size is 4.23.}
\end{figure}

We demonstrate this here in the context of the city of Kolkata, which captures two aspects critical to our thesis --- strong heterogeneity in terms of personal contact networks, as well as areas of great social deprivation, both of which, as will be seen, can lead to the rapid spread of epidemics. While this is a specific choice made by our access to publicly available data [9, 11], its relevance is global. Social deprivation and high population densities among migrant workers in Singapore have recently been held responsible for a second wave of COVID-19 spreading [16, 17], and similar conditions among migrant workers in the UAE are a portent of similar trends. We assert therefore that our highlighting of these issues in the context of Kolkata has global and urgent relevance to the current pandemic.

In the first subsection we focus on the role of heterogeneity, where infections spread via the contact networks of individuals, while in the second, we focus on the role of social deprivation in infection spreading.

\subsection{Profiles of infection spreading via contact networks of individuals}

\begin{figure}
  \includegraphics[width=\columnwidth]{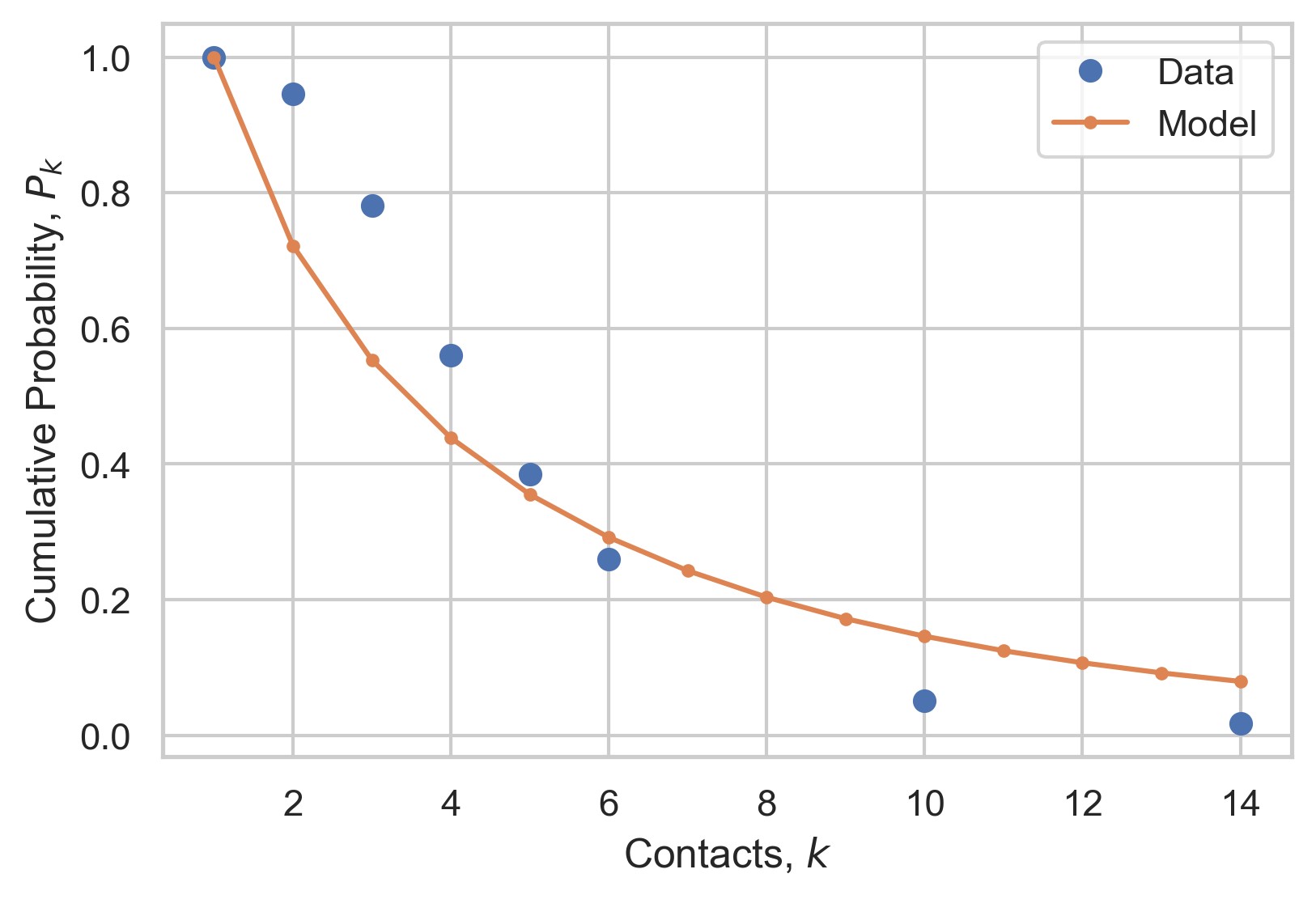}
  \caption{\label{fig-2}
  Cumulative probability distribution of contact networks of Kolkata under strict lockdown. The data (blue dots) yields a mean contact network size of $\langle k \rangle = 4.158$, which is fitted (red line) to the Newman model~\cite{newman-2002} with a mean contact network size of $\langle k \rangle = 4.1568$, yielding parameters $\alpha = 1.0$, $\kappa = 10.44$ and a variance $\langle k^2 \rangle = 45.509$.}
\end{figure}

In most democracies such as India, it would be considered a violation of privacy to have extensive lists of individual contacts made publicly available; additionally, the surveillance required to gather details of where people move and thus whom they might infect in public places, would be even more a violation of democratic rights. We, therefore, focus on conditions of strict lockdown, which are (at least theoretically) valid in India as this paper is being written. Under these conditions, we postulate that the contact network of an individual is limited to contacts within his or her household. For Kolkata, publicly available data [9] leads to the plot in Fig.~\ref{fig-1} of the cumulative probability distribution of the household size.

This leads to the plot in Fig.~\ref{fig-2}, which is the cumulative distribution of the number of contacts a person has (blue dots) which was subsequently fitted (red line) by a power-law/exponential form~\cite{newman-2002}. The fit preserves the mean contact network size of $\langle k \rangle = 4.158$ to a very good approximation, and yields the parameters $\alpha = 1.0$, $\kappa = 10.44$, which will be used in the analysis that follows.

\begin{figure}
  \includegraphics[width=\columnwidth]{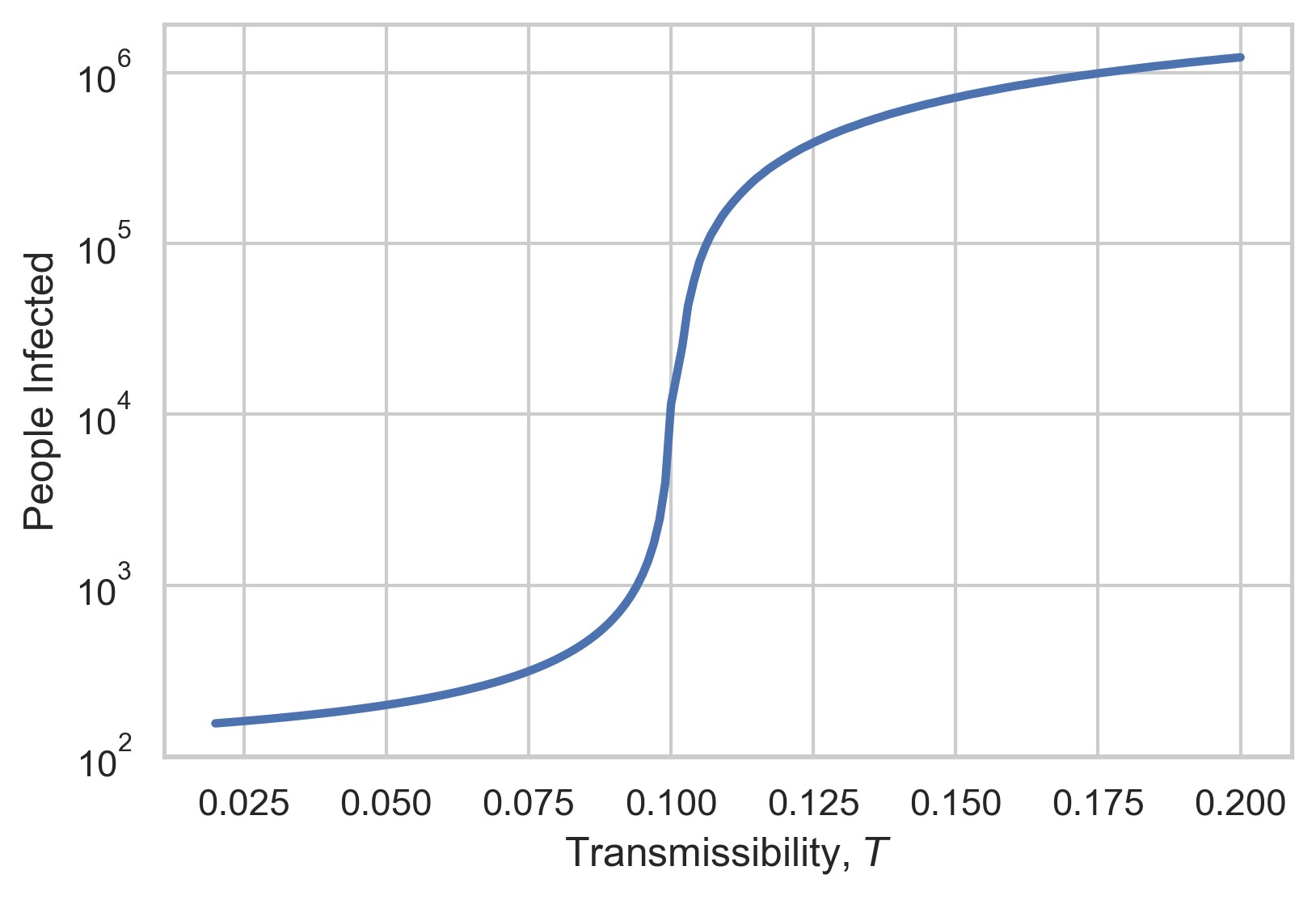}
  \caption{\label{fig-3}
  The number of people infected for the city of Kolkata with a population of $4,261,627$, as a function of transmissibility $T$. A transition to an epidemic happens at a critical value of $T$, for $T_\text{c} = 0.1005$.}
\end{figure}

Using the analysis of the previous section, Eq. (4), we obtain the curve of infected people vs transmissibility shown in Fig.~\ref{fig-3}. The prediction for a transition to an epidemic is at a critical value of $T_\text{c} = 0.1005$, which is much lower than $T_\text{c} = 0.3167$ for a homogeneous model with the same $\langle k \rangle$; the homogeneous model thus clearly underestimates the risk of the epidemic here and in general.

\begin{figure}[b]
  \includegraphics[width=\columnwidth]{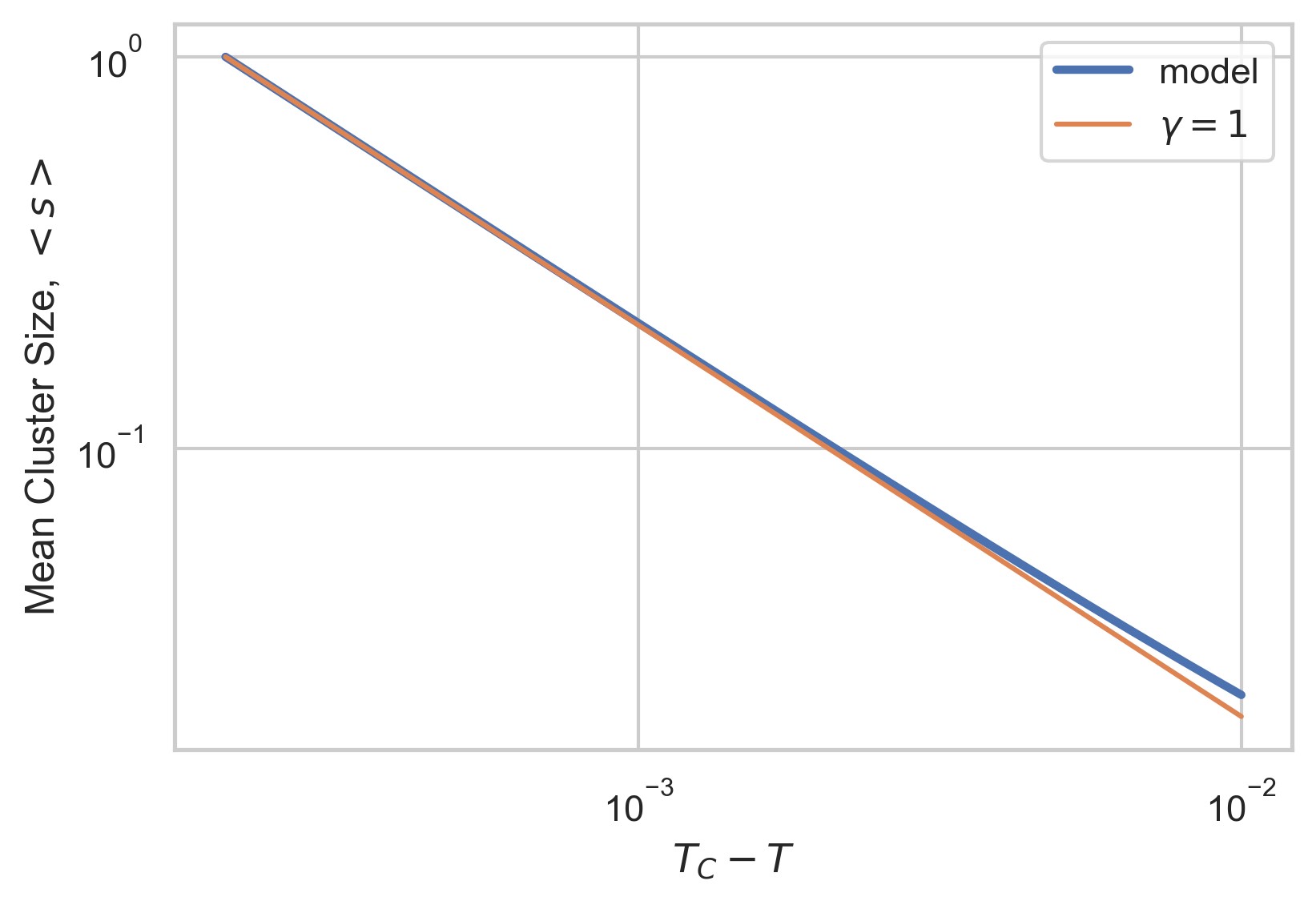}
  \caption{\label{fig-4}
  The mean cluster size $\langle s \rangle$ of infected people against $T_\text{c}- T$ (blue line) fitted to a value of $\gamma = 1$ (red line).}
\end{figure}

The mean cluster size of infected people diverges with an exponent $\gamma$ of 1 (see Eq. 5), as shown in Fig.~\ref{fig-4}.

\begin{figure*}
  \begin{tabular}{cc}
    \includegraphics[width=0.45\textwidth]{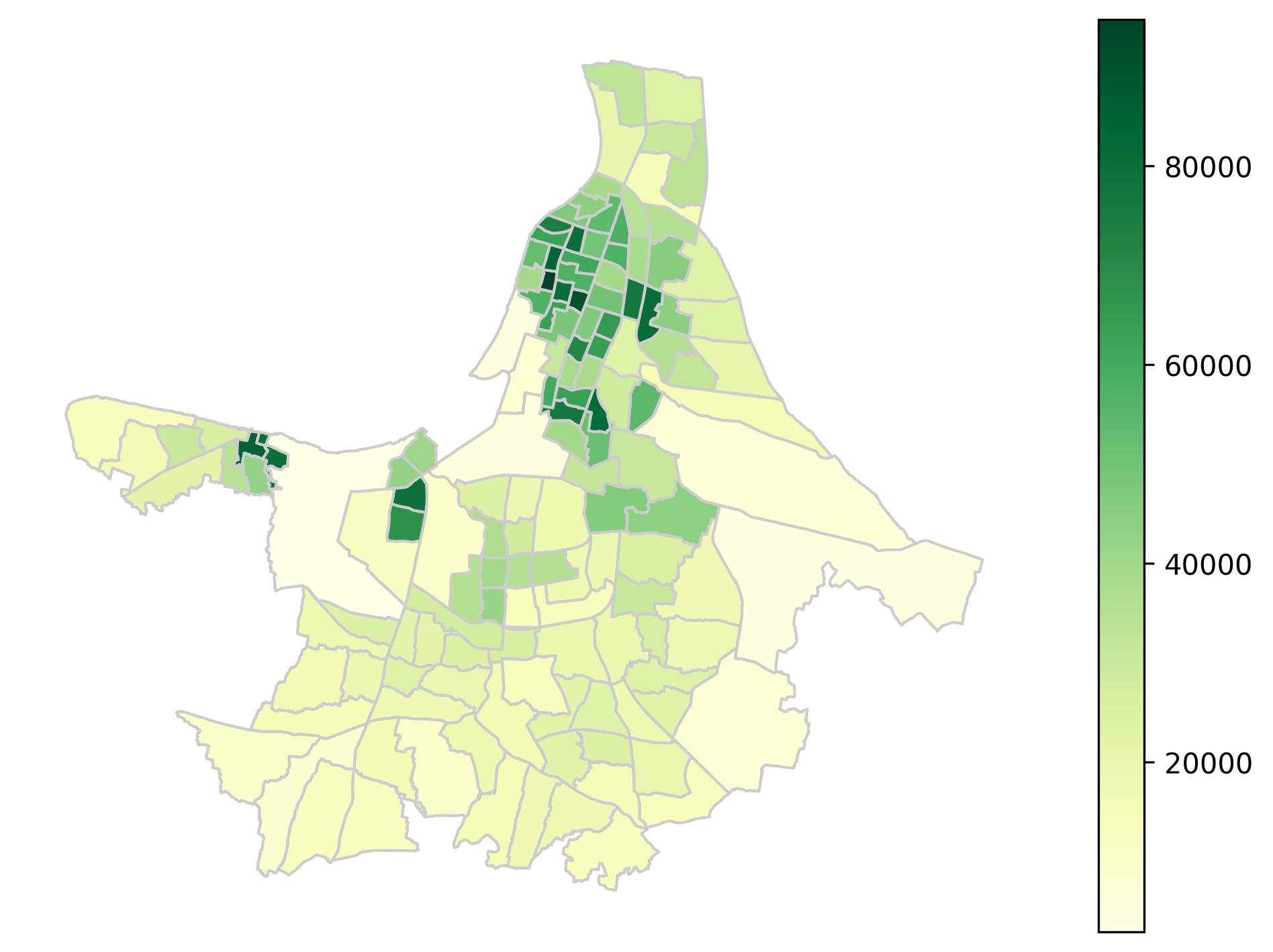} &
    \includegraphics[width=0.45\textwidth]{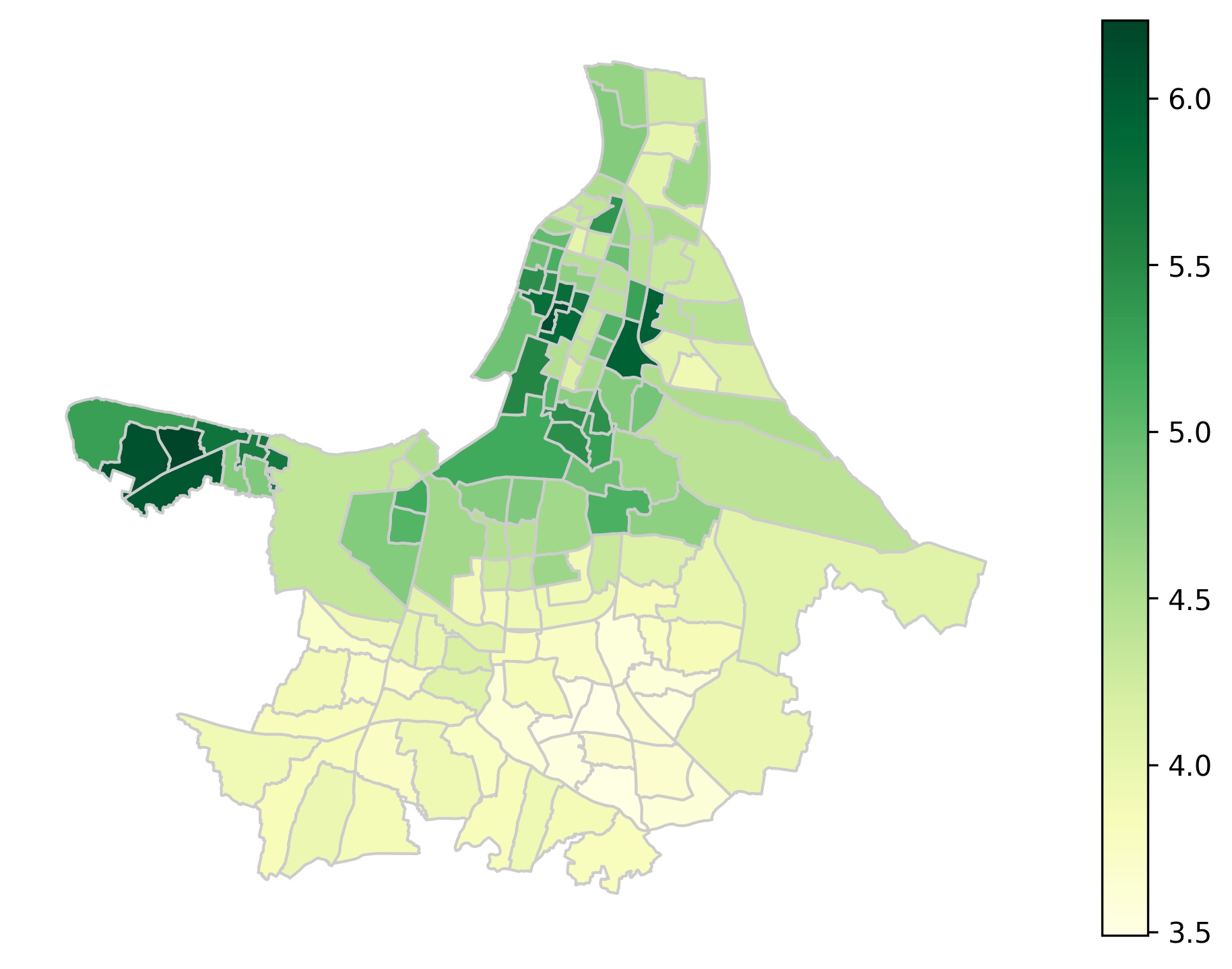} \\
  \end{tabular}
  \caption{\label{fig-5} (A) The population density of the city of Kolkata measured by no. of persons/square kilometre, as a function of the city wards which are delineated above. The legend alongside indicates the colour coding in terms of population density. (B) The household size distribution across different wards in Kolkata. The legend to the right is a colour-coded key encoding household size, so that, e.g. household sizes of 6 and over are indicated by the darkest shades of green.}
\end{figure*}

\begin{figure}[b]
  \includegraphics[width=\columnwidth]{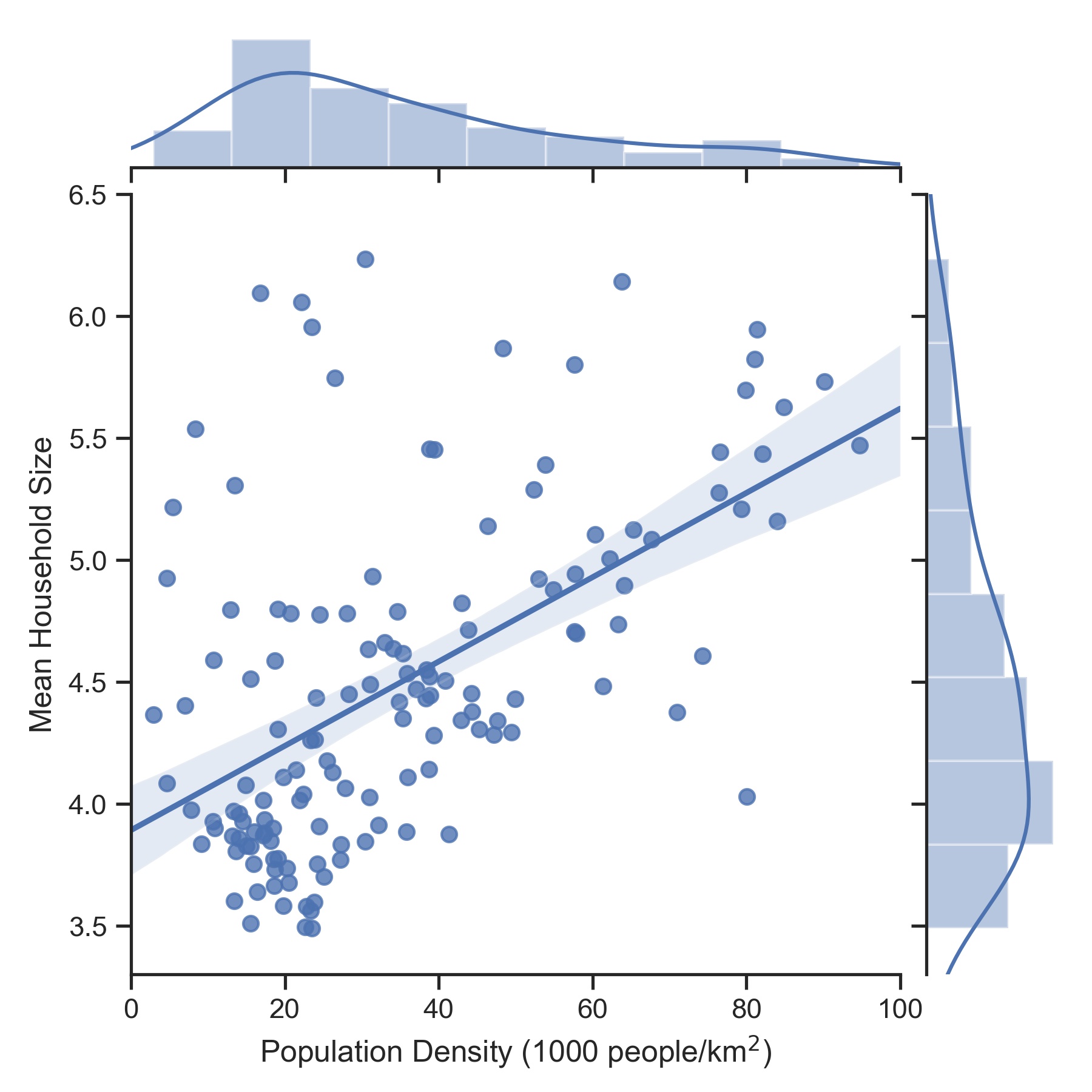}
  \caption{\label{fig-6} A linear scatter plot of mean household size vs. population density in Kolkata. The histogram of population density is shown on top of the main scatter plot, while that for mean household size is on the right of the main scatter plot.}
\end{figure}

The above city-wide prescription is strongly modified when we look at granularity at the ward-level --- a ward in Kolkata is defined as a locality for which census data are available [9] in terms of various demographics. Wards are a first indicator of heterogeneity since some are more densely populated than others, as shown in Fig.~\ref{fig-5}A. The household size distribution per ward, which is independently available from the data, is given in Fig.~\ref{fig-5}B. As we might expect, there is a reasonable correlation between the two (Fig.~\ref{fig-6}), i.e. areas where there is a high density of population are also those where the household sizes are large. These seem to be clustered to the north and the west of the city (Figs.~\ref{fig-5}A and \ref{fig-5}B).

\begin{figure}[b]
  \includegraphics[width=\columnwidth]{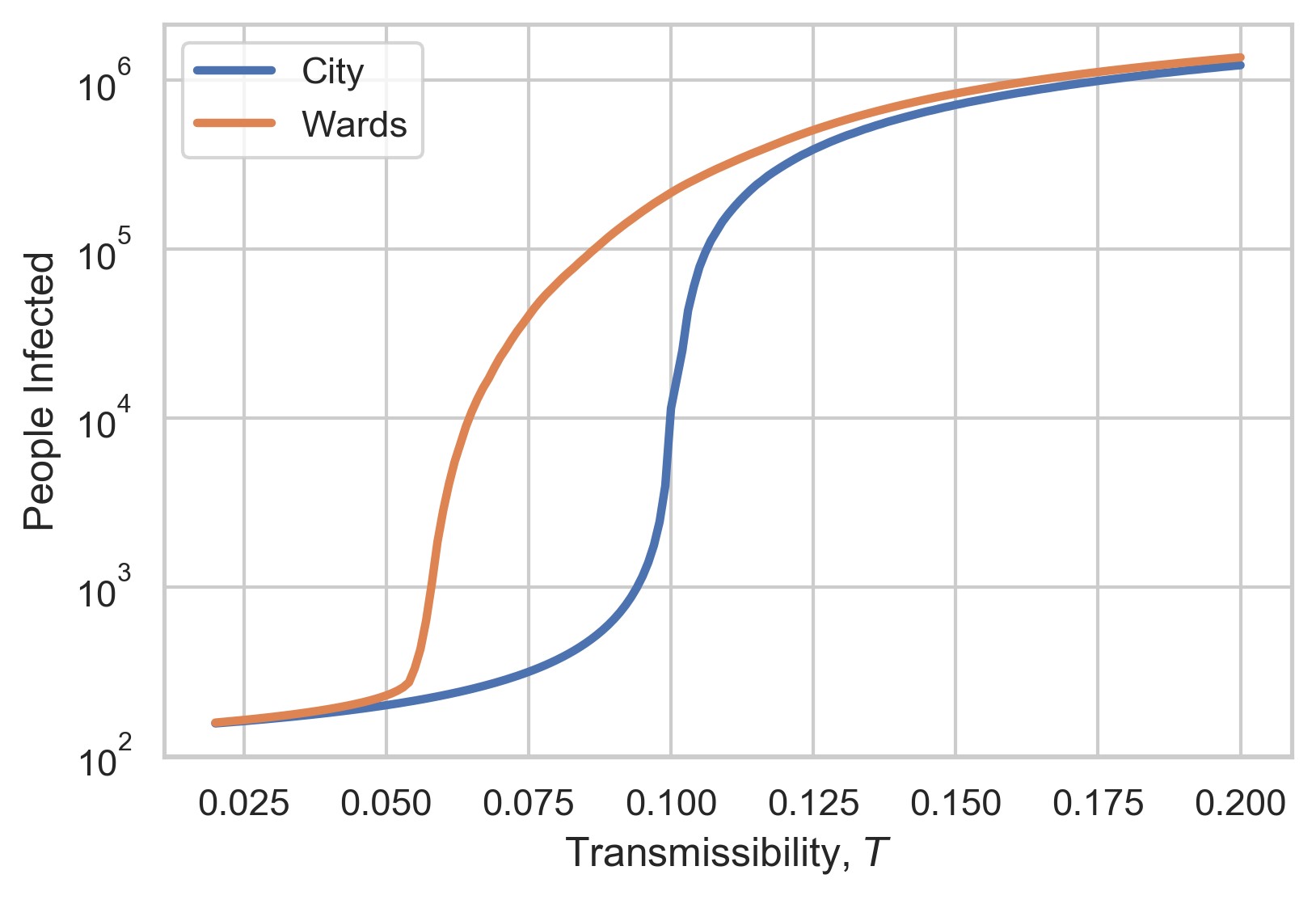}
  \caption{\label{fig-7} The number of people infected as a function of transmissibility $T$ using a) a ward-based picture (orange line) and b) a city-based picture (blue line). The granularity of the ward-based picture results in strong heterogeneities of contacts, which allow for a faster spreading of the epidemic than if one assumes the more macroscopic picture of contact networks following the same degree distributions across the city. While the city-based picture corresponds to a single $T_\text{c} = 0.1005$ (Fig.~\ref{fig-3}), the ward-based picture allows a distribution of $T_\text{c}$ from $0.055$ to $0.138$.}
\end{figure}

The heterogeneity introduced by wards in terms of their household sizes introduces important heterogeneities into the contact networks of the individuals in them. Using the formalism above, this in turn introduces important differences in the number of individuals infected, since the populations of the wards have distinct characteristics. As a consequence, the transition to an epidemic sets in much faster if we use this level of description, as is seen in Fig. \ref{fig-7}. Since, under strict lockdown, it is reasonable to assume that people will interact more within their wards than city-wide, we believe that the orange curve is a better representation for the spread of the epidemic than the blue (city-wide) curve.

\begin{figure}
  \includegraphics[width=\columnwidth]{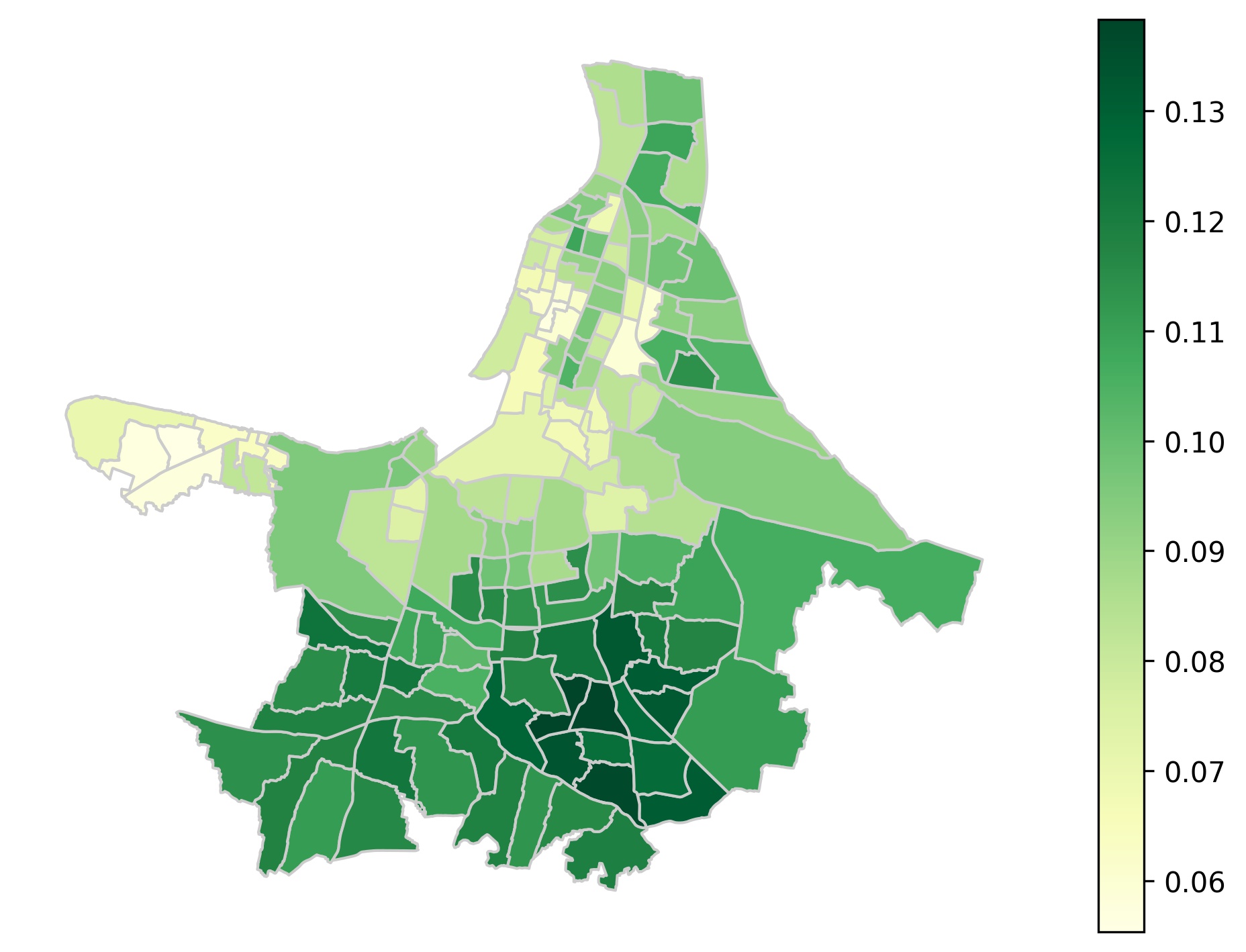}
  \caption{\label{fig-8} The distribution of critical transmissibility $T_\text{c}$, ward-by-ward, across the city of Kolkata. Note that the higher the $T_\text{c}$, the slower the spread of infection. The most vulnerable areas (lightest colours) in terms of epidemic spread are to the north and west of the city.}
\end{figure}

Another way of seeing this is to say that if we assume that the wards are self-contained, each one is associated with a specific critical transmissibility $T_\text{c}$ (Fig.~\ref{fig-8}). The epidemic spreads as soon as $T_\text{c}$ is attained within a ward, with a corresponding explosion in the number of people infected. Note that the higher the $T_\text{c}$, the slower the spread of infection. The most vulnerable areas in terms of epidemic spread are to the north and west of the city (lightest colours), which correlate well with the areas of high population density and household size shown in Figs.~\ref{fig-5}A and \ref{fig-5}B.

\begin{figure*}
  \begin{tabular}{cc}
    \includegraphics[width=0.47\textwidth]{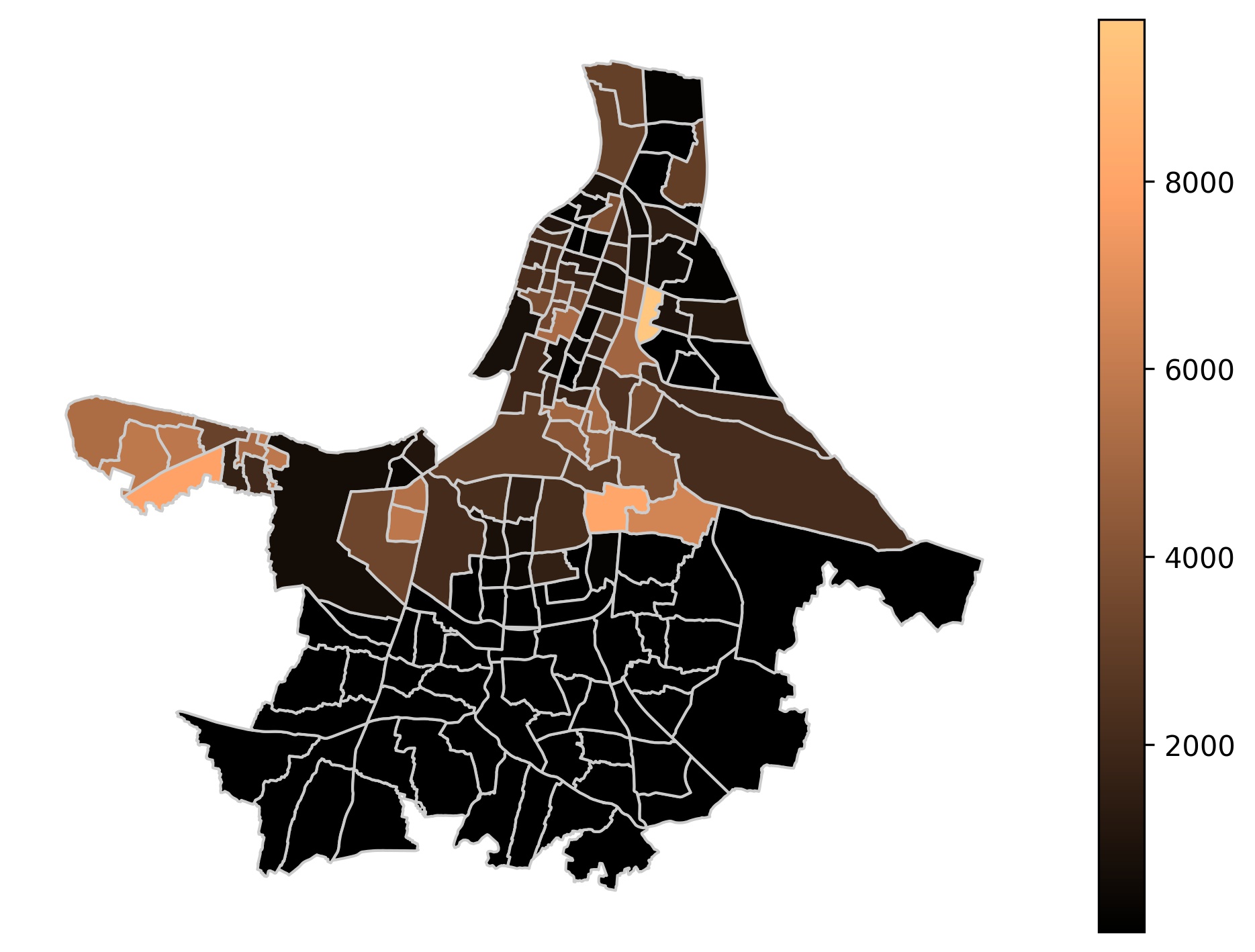} &
    \includegraphics[width=0.47\textwidth]{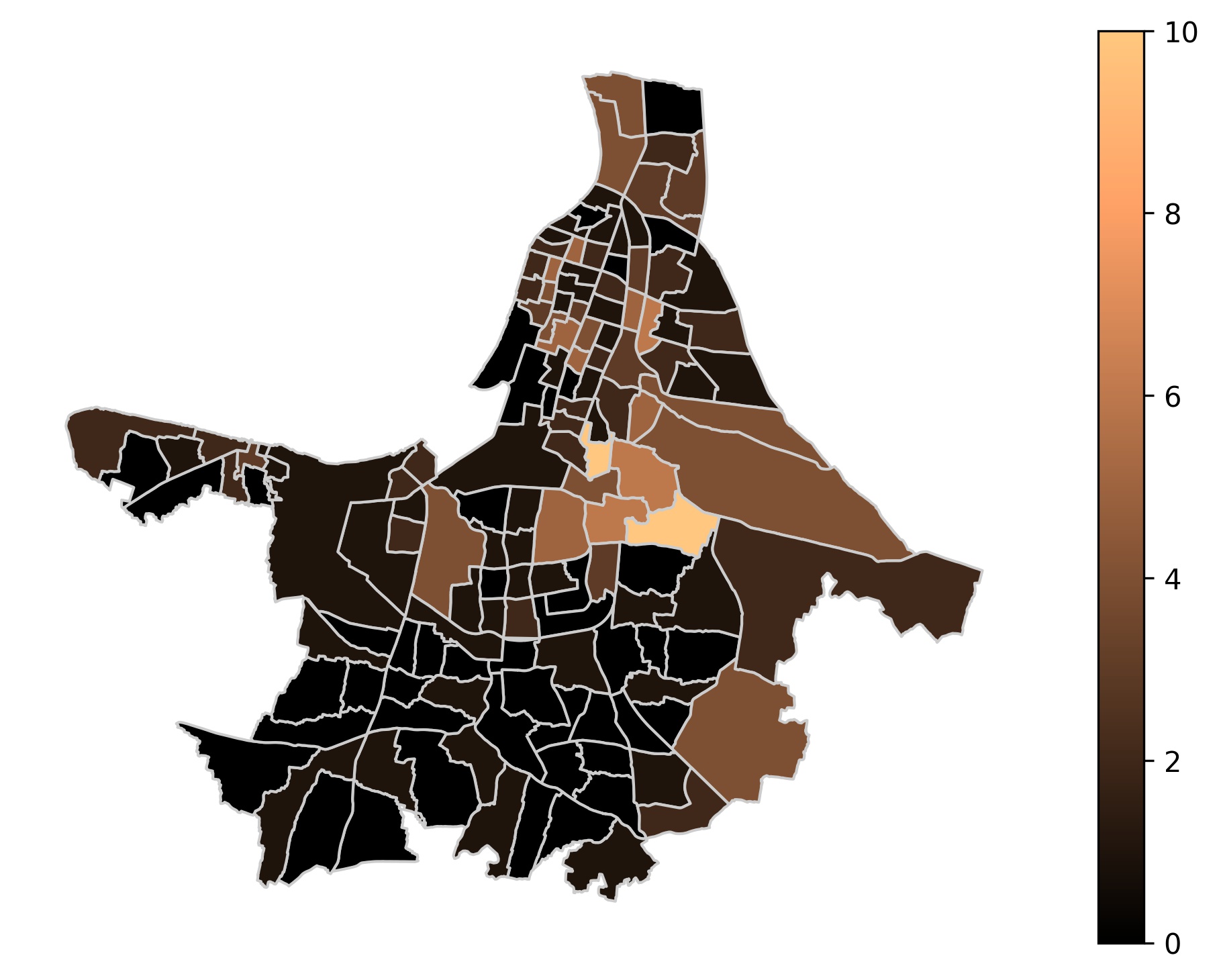} \\
  \end{tabular}
  \caption{\label{fig-9} (A) Ward-based numbers of infected people for the city of Kolkata for transmissibility $T = 0.1$, similar to the $T_\text{c}$ for the entire city. Note that the infections that are localized to the vulnerable areas of the north, centre, and west of the city gradually include the south for higher values of $T$.~\cite{majumdar-2020}  (B) Number of containment zones per ward in the city of Kolkata as of April 28, 2020 [14].}
\end{figure*}

We have designed a web application [18] to demonstrate the spread of the epidemic. Fig.~\ref{fig-9}A demonstrates this for a single value of the transmissibility $T = 0.10$ which is the threshold for the city taken as a whole. The heterogeneous nature of outbreaks is apparent, as some vulnerable wards (mainly to the north, centre and west) have an infected population numbering several thousands, while others (mainly in the south) are yet to see any significant infections at all. As $T$ increases (e.g. by easing the lockdown and increasing the frequency of contacts between people), more wards would cross their local critical threshold and begin to see significant infections. While some of the hotspots predicted in Fig.~\ref{fig-9}A have not yet been empirically observed, we assert that these are areas of potential risk.

We compare these results of our model to the situation in the city as of late April 2020. Since the locations of infectious clusters have not yet been published, we look at the number and locations of those regions which have been designated as containment zones. Containment zones are set up by the government to contain the disease within a defined geographical area (usually a city block) following multiple confirmed infections in the area, with a view to breaking the chain of transmission and preventing the spread of the infection to new areas. Each zone is geographically quarantined with enhanced active surveillance [19]. As on April 28, 2020, out of 287 such containment zones in the city of Kolkata, 227 have been identified by ward [19]. Figure~\ref{fig-9}B shows the number of such zones within each ward. Although most containment zones in Fig.~\ref{fig-9}B are geographically aligned with the areas we have identified as being at risk (Fig.~\ref{fig-9}A), there are a few areas of local outbreaks which our model did not predict in the south-east of the city. This area was urbanized relatively recently and became densely populated in the period after 2011, the date of the last census [9], which is the source of our data for this part of our research.

Before leaving this subsection, we summarise the nature of our findings. Heterogeneity of contact networks plays a crucial role in the transmission of disease, even if we take a macroscopic viewpoint on the population of a city like Kolkata. When we take a look at individual wards and use available data to construct more realistic contact networks of individuals, we note that areas of high population density are strongly correlated with large household sizes, and so, within our present approximation, with extended contact networks. These are in turn of crucial importance in the spread of disease, as our predictions demonstrate, predictions which in fact are well correlated (Fig.~\ref{fig-9}) with existing governmental preventive approaches [19].

In the next subsection, we will focus on areas of high social deprivation. These could, in general, be migrant housing in Singapore or Dubai, but in the present instance, are based on data on the slums in Kolkata. It will be seen that such areas are particularly vulnerable to becoming hotspots for disease transmission.

\subsection{The role of social deprivation in disease transmission --- a case study of slums}

We have used existing data [11] on slums in Kolkata as well as publicly available census data [9] for our analysis in this section. In addition to dismal living conditions wherever overcrowding is the norm (and a major mechanism for the forced enhancement of human contacts), a major indicator of social deprivation is the lack of literacy that usually obtains in slums. The latter is particularly important in the context of COVID-19 spreading since it translates into a lack of awareness for the very necessary preventive measures at an individual and collective level that would help fight the virus.

\begin{figure*}
  \begin{tabular}{cc}
    \includegraphics[width=0.47\textwidth]{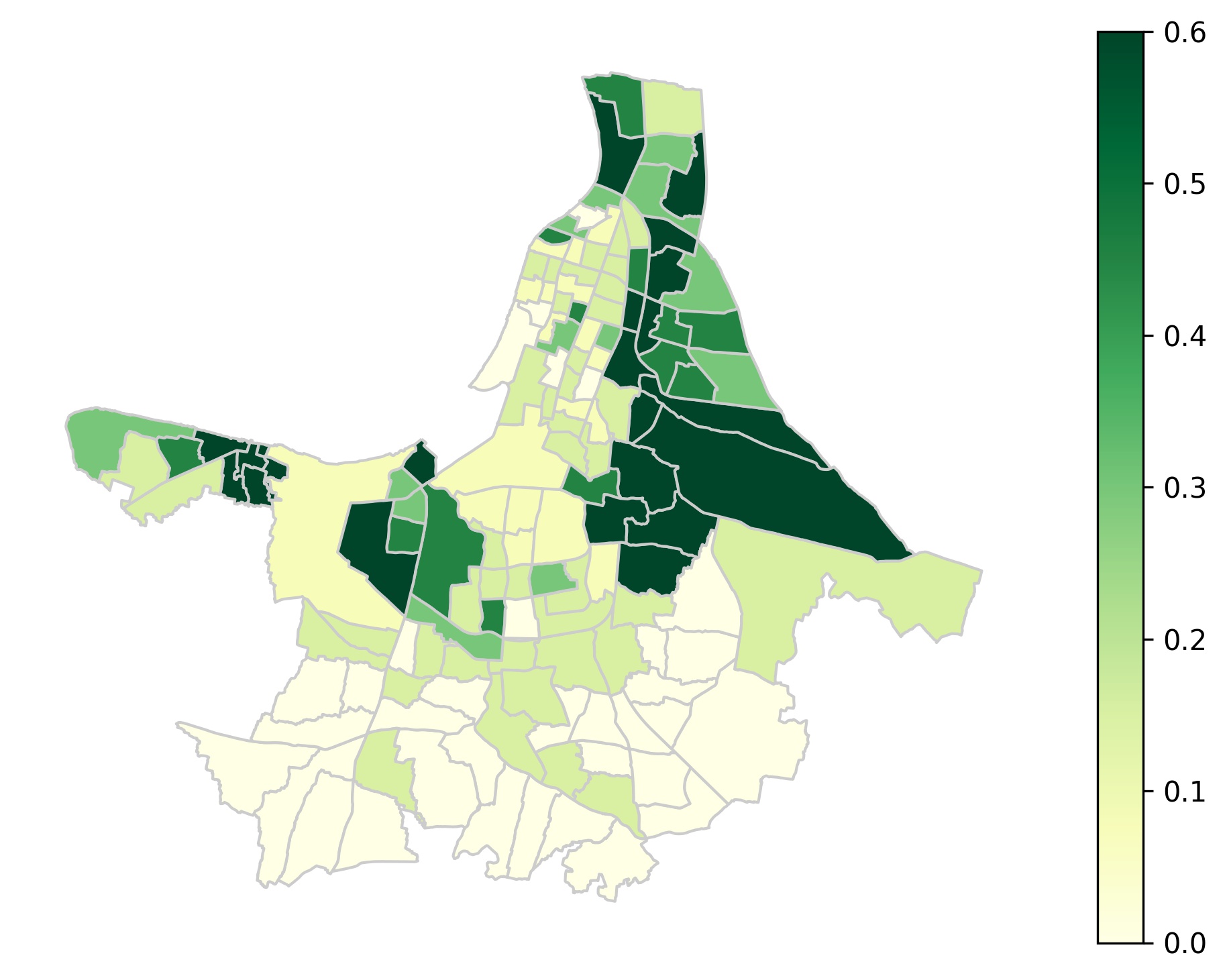} &
    \includegraphics[width=0.47\textwidth]{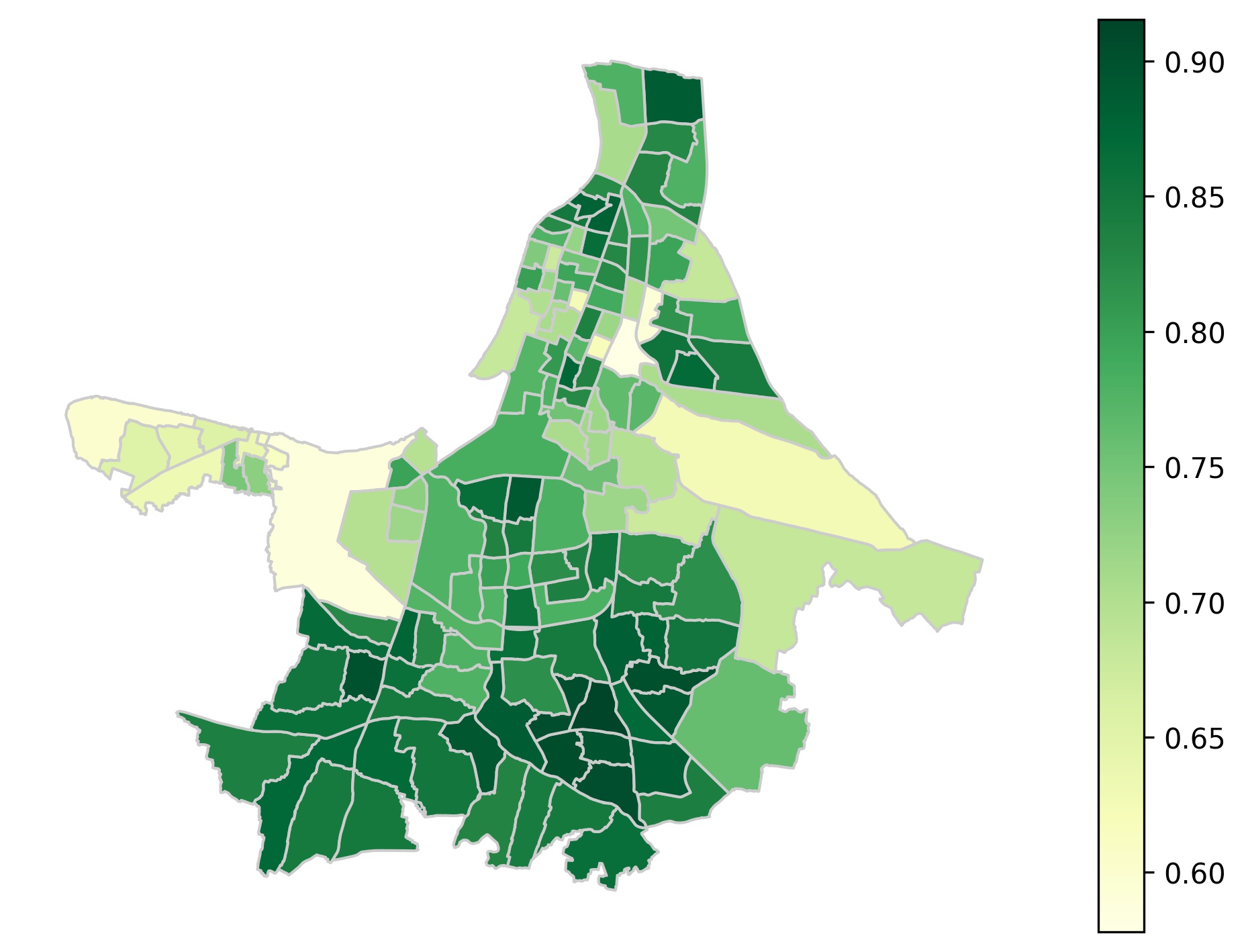} \\
  \end{tabular}
  \caption{\label{fig-10} The distribution of the (A) fraction of population living in slums, and (B) fraction of literate individuals, ward-by-ward, across the city of Kolkata.}
\end{figure*}

We first use the census data to look at the lack of literacy in the slum population. Fig.~\ref{fig-10}A shows the ward-by-ward fraction of slum dwellers in Kolkata; this appears to be the complement of Fig.~\ref{fig-10}B, which shows the fraction of literate people computed ward-by-ward for the city.

\begin{figure*}
  \begin{tabular}{cc}
    \includegraphics[width=0.47\textwidth]{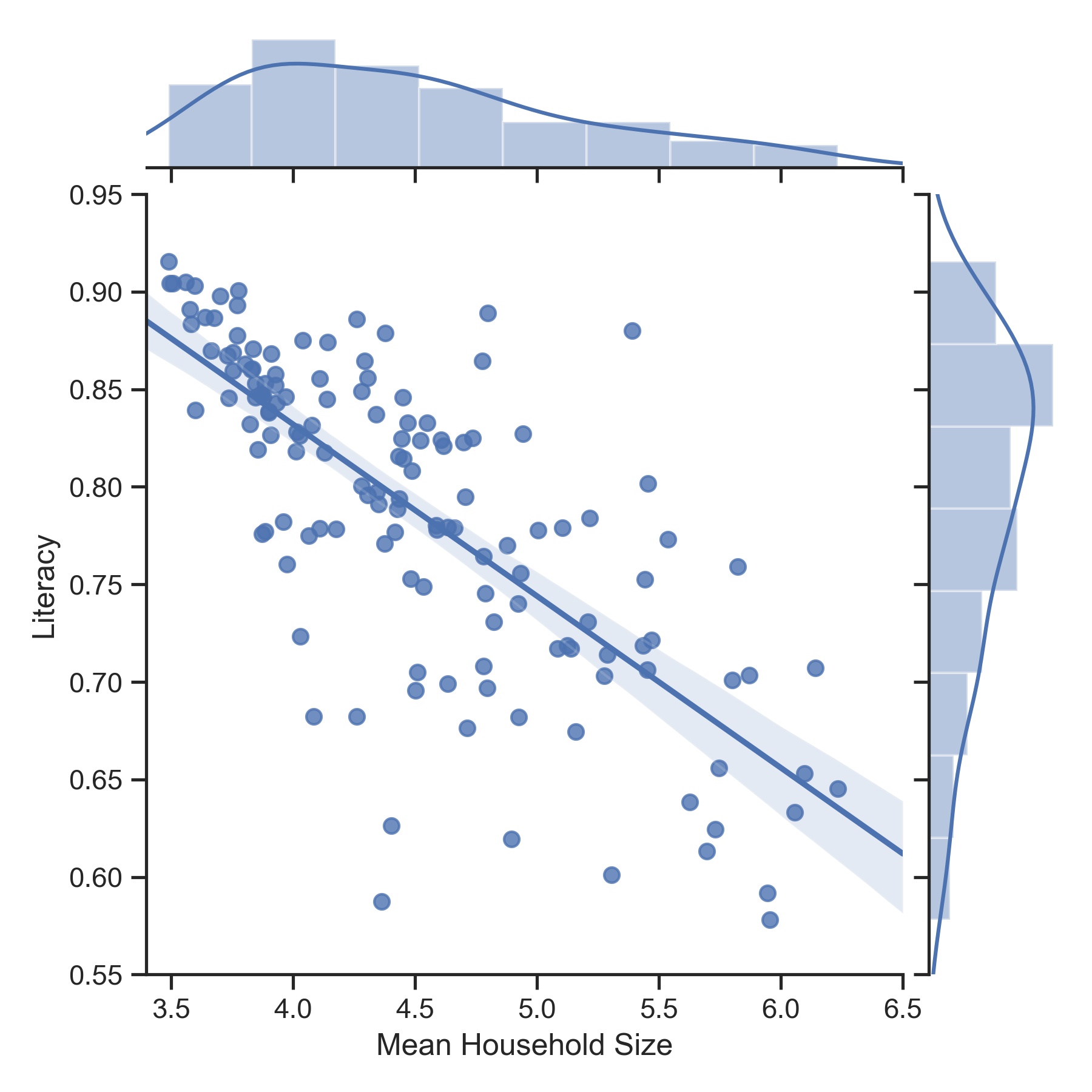} &
    \includegraphics[width=0.47\textwidth]{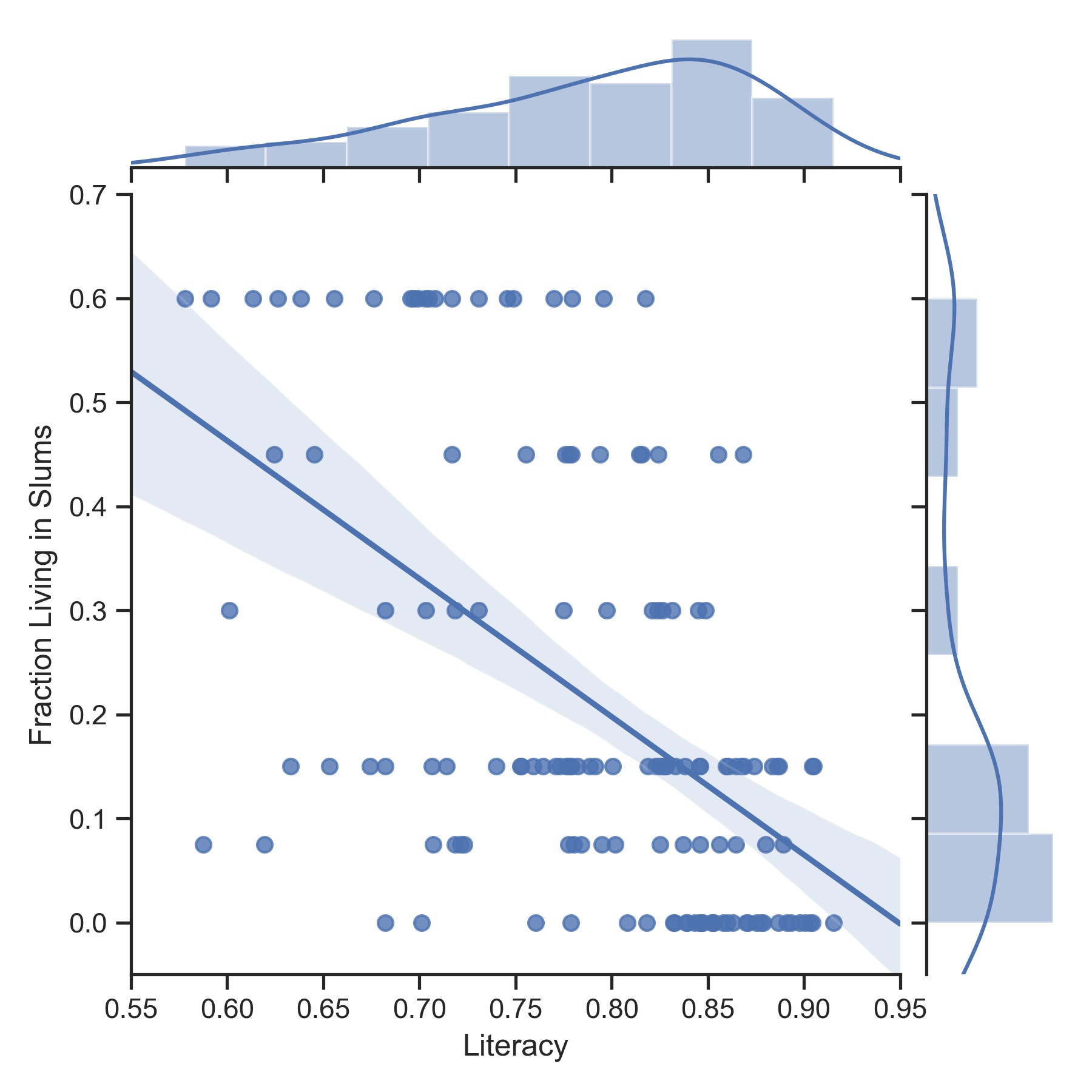} \\
  \end{tabular}
  \caption{\label{fig-11} The distribution of the (A) fraction of population living in slums, and (B) fraction of literate individuals, ward-by-ward, across the city of Kolkata.}
\end{figure*}

In Fig.~\ref{fig-11}, we quantify the above picture by scatter plots showing the correlation between (A) literacy and household size, and (B) the fraction of population living in slums and literacy. All the data clearly show there is a strong negative correlation between literacy and household size, and that in slums, in particular, literacy rates tend to be low.

\begin{figure*}
  \begin{tabular}{cc}
    \includegraphics[width=0.47\textwidth]{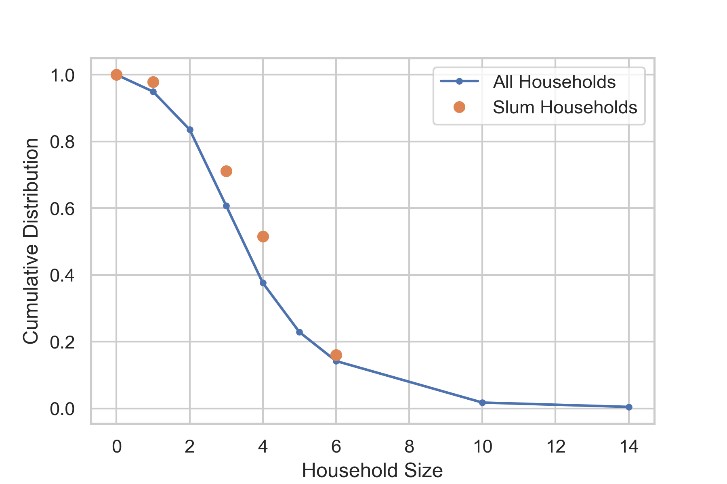} &
    \includegraphics[width=0.47\textwidth]{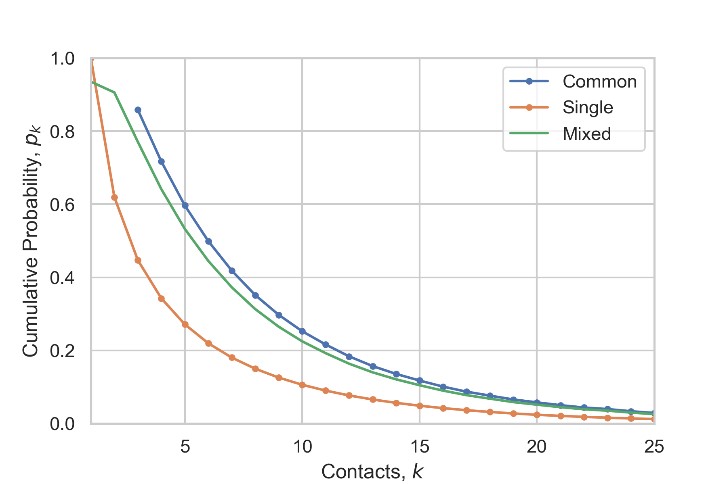} \\
  \end{tabular}
  \caption{\label{fig-12} (A) Plot of the cumulative probability distribution for household sizes for all (blue line) and slums (red dots). (B) Plot of the cumulative probability distribution for contacts for households with private toilets (red line), for those who have to share common facilities (blue line) and for the resultant (green line), based on a percentage ratio 83:17.}
\end{figure*}

We now use recently collected data specific to slums in Kolkata [11] that, in addition to providing a household size distribution for them, also provides a measure of deprivation, in this case, due to overcrowding. Many slum families share toilet facilities and often depend on public borewells to get their water. From the point of view of our research, this, even under strict lockdown, forcibly extends their contact networks. In Fig.~\ref{fig-12}A we show that the household size distribution for slums [11] is only slightly larger overall than that for the overall population computed based on census data [9]. However, the fact that 83\% of the slum population in Kolkata share toilet facilities as opposed to 17\% who have private toilets, causes a dramatic change in the degree distribution of contacts as shown in Fig.~\ref{fig-12}B. The orange curve corresponds to the degree distribution for slum dwellers with private toilet facilities, while the blue one corresponds to that where facilities are shared with at least one other family, so that with the percentages mentioned above, the resulting degree distribution is given by the green curve.

\begin{figure*}
  \begin{tabular}{cc}
    \includegraphics[width=0.47\textwidth]{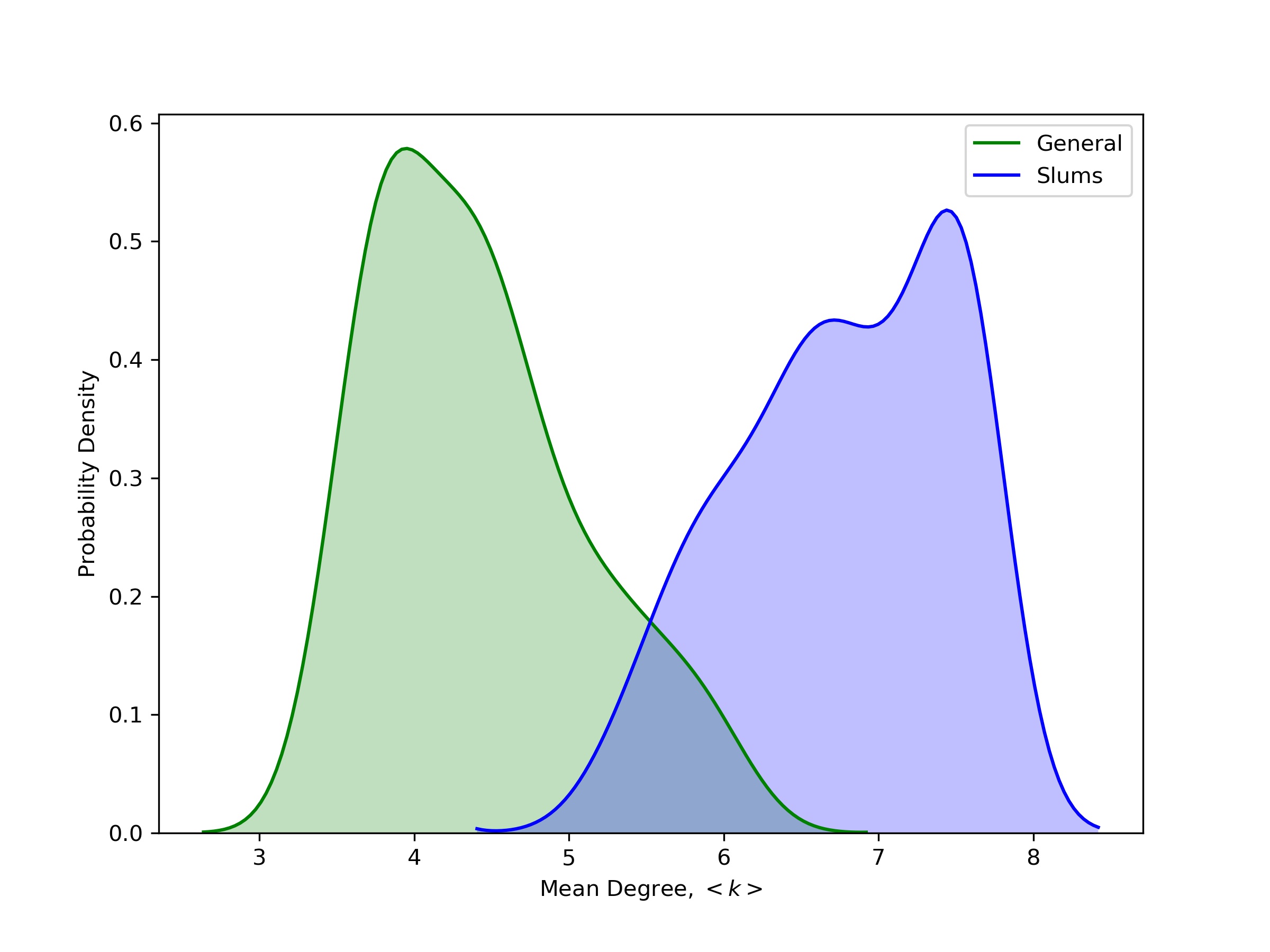} &
    \includegraphics[width=0.47\textwidth]{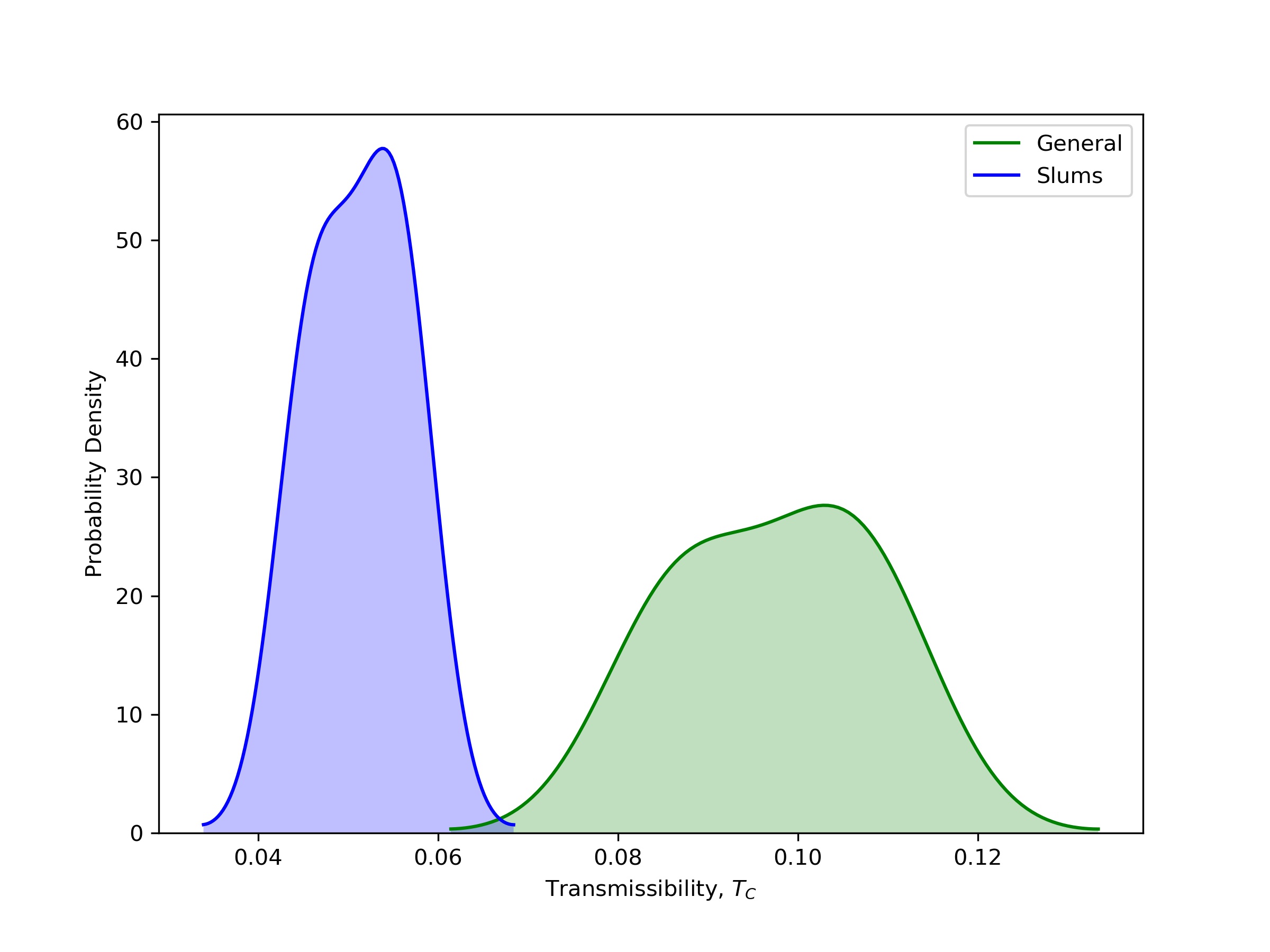} \\
  \end{tabular}
  \caption{\label{fig-13} The probability density function of ward-by-ward (A) mean degrees, and (B) transmissibility Tc, for the general (green) and the slum populations (blue) of Kolkata. Mean degrees are shifted to the right for the slum population [9, 13]. The Tc curves indicate that the critical transmissibility Tc sets in much earlier for the slum population than for the general population.}
\end{figure*}

As we will see, this sharing of facilities has a dramatic effect first, on the degree distribution (Fig.~\ref{fig-13}A), and next, on the critical transmissibility $T_\text{c}$ (Fig.~\ref{fig-13}B) for the slums. For both Figs.~\ref{fig-13}A and \ref{fig-13}B, the green distribution represents the entire population, while the blue one is specifically for the slums. We note that the sharing of facilities such as toilets leads to a large effective shift in the mean degree distribution for the slums vis-\`a-vis that of the general population, cf. the shift of the blue curve from the green curve in Fig.~\ref{fig-13}A --- despite our rather conservative estimate of this (see the Discussion for further details). This has an even more dramatic effect on the critical transmissibility $T_\text{c}$ for slums, as will be seen in Fig.~\ref{fig-13}B. The sharp peak for the slum population sets in at a much lower value of $T_\text{c}$, so that the slum population are much more vulnerable to epidemics than the general population taken as a whole.

\begin{figure*}
  \begin{tabular}{cc}
    \includegraphics[width=0.47\textwidth]{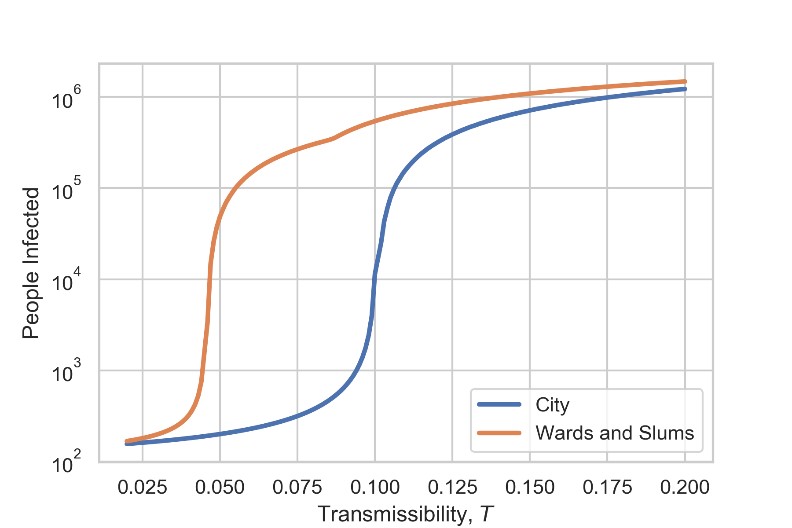} &
    \includegraphics[width=0.47\textwidth]{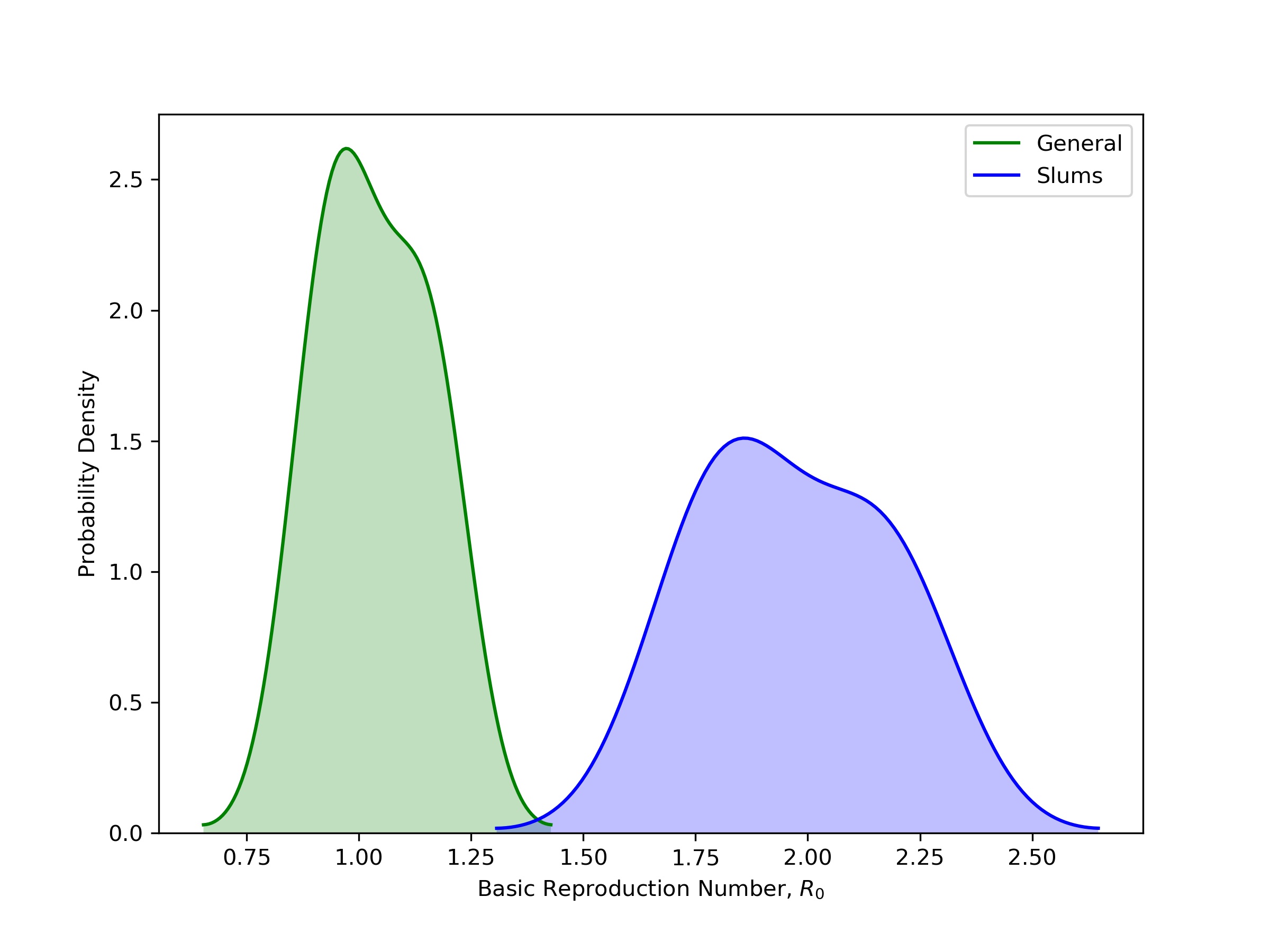} \\
  \end{tabular}
  \caption{\label{fig-14} (A) Comparison of city-wide transmissibility curve which shows a transition at 0.1005 with the ward-by-ward transmissibility curve, now including the effect of slums, which shows that the transition to an epidemic occurs even earlier than what is predicted by the ward-by-ward graphs in Fig.~\ref{fig-7}. (B) The probability density function of the ward-by-ward basic reproduction number $R_0$ for the general (green) and slum (blue) populations in the city.}
\end{figure*}

In Fig.~\ref{fig-14}A, we take into account the effect of the slum population obtained from the data of [11] to compute the number of people infected ward by ward (orange curve), to be compared with the city-wide computation (blue curve). We notice a much sharper increase in the number of people infected, relative even to the ward-by-ward graph of Fig.~\ref{fig-7}. As a result, the effective mean reproduction number $R_0$, obtained from Eq. (6) is greater for the slum population than the general population for the same value of $T_\text{c}$, as will be seen from Fig.~\ref{fig-14}B.

\section{Discussion}

\subsection{Key findings of our study, in the context of past approaches}

Most approaches to date on the COVID-19 spread have involved homogeneous mixing (see e.g. [15]), whereas ours is based on heterogeneous mixing, depending on the degree of the nodes of the underlying contact network. The main effect of this heterogeneity is to reduce the threshold to the epidemic, so that the epidemic spreads faster than it would with homogeneous mixing.

Our aims in this paper are, therefore: first, to insert the heterogeneity of contact networks in theoretical approaches to model the spread of COVID-19 and second, to use this formalism to provide an insight into the role of social deprivation.  We have looked at slums as a source of high-degree connectivity, which, combined with low economic development (e.g. the anti-correlation with literacy) provide explosive ingredients for a transition to an epidemic. Both of these ingredients have the effect of considerably advancing this transition in the areas where they are prevalent, and hence to the entire population of the city of which they are part.  While our data analysis is specific to Kolkata, it is very much more general in its applicability to cities with areas of high connectivity and social deprivation, of which the recent example of Singapore [16, 17] is only one example [20].

Our approach is also able to identify hotspots on the basis of individual contact networks in a scientific and objective way, thus allowing for an impartial way of identifying areas where containment should be enforced as an overall means of prevention. Our predictions are in good agreement with empirically obtained government data [19], where the latter identify containment zones as areas where infections have already occurred. On the other hand, our approach goes further, allowing for such identification even before infections have occurred, based on household sizes and individual contact networks obtained therefrom; this would form a good basis for preventive measures.

\subsection{Potential Limitations}

Without access to data on individual contact networks, we are limited to household sizes as measures of degree distributions, which they are only under (difficult to implement) conditions of strict lockdown. From this point of view, the level of agreement between our predictions and empirical data on containment zones is indeed remarkable. However, we assert that our results should be interpreted qualitatively rather than quantitatively as a means of highlighting the difference between predictions arising from heterogeneous and homogeneous contact networks; for example, our predictions for $T_\text{c}$ show clearly that contact network heterogeneity causes infections to spread much more rapidly than might be imagined on the basis of homogeneous theories.

Another limitation arises from the availability of data on slums. In [11], 800 slums in Kolkata were studied via sampling methods, without any information on their exact location. We have therefore had to study the slums collectively, without any information on where exactly, and how extensive, they are in any particular area. The best that we have been able to do is, via census statistics on the percentage of slums per ward [9], compute the ward-based slum statistics to give a qualitative estimate of the ward-based $T_\text{c}$. Obviously, the approximation involved in doing this, i.e. treating the slums per ward as a contiguous unit when they are in fact distributed, is considerable, but in the absence of data, this is an approximation that we are forced to make. Another approximation involves the way in which we compute the revised degree distribution for slums due to shared toilet facilities. In the absence of spatial information on how many families share these facilities, we have taken a conservative approach, assuming that a family of size $n$ shares facilities with only one other family of size $n_1$, and using this idea to construct the relevant distributions. Of course in actuality, many families may well share the same toilet, (or indeed the same access to drinking water via public borewells, which we have not even considered), so we are, if anything, underestimating the severity of the resulting overcrowding and epidemic spread. Another point in this context concerns what we mean by the ``general'' population. The census data [9] that we have based this on includes people of all classes, including slum dwellers, in the city of Kolkata; the specific data on slums [11] is both more recent and focuses only on the 800 slums considered in Kolkata. We reiterate therefore that, as mentioned above, our estimates should be taken as a conservative qualitative indicator of the effects of social deprivation, rather than an accurate quantitative estimate. On a more positive note, our methods can be applied, if the requisite data become available in a rigorous and detailed format, to areas of social deprivation worldwide with a view to estimating their vulnerability to epidemics. Last but not least, while our predictions on hotspots in Fig.~\ref{fig-9}A are quite well correlated with the empirical data on containment clusters in Fig.~\ref{fig-9}B, we mention here that our predictions are based on 2011 census data, when some of the areas in the south-east of the city were far less developed and populated than they now are. Had we had access to recent data, this new demography would have been reflected in revised contact network data, and consequent vulnerability to hotspots. 

\subsection{Integrations into current understanding of problem}

There are several ways in which our approach will integrate into current developments. First, there is an emphasis in many countries such as India on containment zones in hotspots --- this is done by looking in hindsight at available data and deciding that areas should be declared as containment zones based on existing infection and death rates, rather than by using network theory. On the other hand, our network-based theoretical approach is predictive; i.e. we can, on the basis of degree-based data, predict where hotspots could occur, and take a preventive rather than curative approach. Also, and importantly, our approach provides a non-controversial way of suggesting hotspots, without touching on sensitive ethnic or religious characterisations.

Another way is for our methods to be used in conjunction with contact tracing apps that are being developed in several countries. An issue that some of these have is that the identification of contacts is done via telephonic links, rather than individual and intimate contact, which is typically how infections spread. If, for example, people in hotspot areas and their contacts within the same neighbourhood are identified, this would be a way of extending the contact network from beyond the household to a larger range; this would reflect conditions where lockdowns are gradually relaxed so that people can move around within a given radius of where they live.

Most importantly, policy-makers in several countries are implementing many empirical measures relating to the gradual relaxation of lockdown, at the time of writing of this paper. These usually involve maintaining strict lockdown in areas that continue to be at risk, while relaxing restrictions in those where there have been few, or no, infections within the recent past. Our formalism, with its detailed predictions of levels of risk in the component areas of a large city, would be invaluable in assisting this process by identifying regions that could be, or should not be, opened up. This is of special importance as economic contingencies compel the world to exit strict lockdowns where possible, while public health contingencies demand that the COVID-19 epidemic is contained to the extent possible.

\subsection{Future directions}

From a scientific point of view, we would like to look at the temporal evolution of the COVID-19 epidemic, in different national contexts, from a network-based point of view. More generally, given the threat that humankind faces from this pandemic, we would welcome collaborations with governmental or other agencies in the design of realistic counter-measures.
 
\section*{Declaration}
The authors declare that the research was conducted in the absence of any commercial or financial relationships that could be construed as a potential conflict of interest.

\acknowledgments{
  Anita Mehta is grateful to the Max Planck Institute for Discrete Mathematics, Leipzig and the Centre for Linguistics and Philology, University of Oxford for their support.
}

\end{document}